\documentclass{ws-ijmpcs}

\begin{document}

\markboth{Boito et al.}
{Low-energy constants and condensates from the $V-A$ spectrum}

%
\catchline{}{}{}{}{}
%

\title{LOW-ENERGY CONSTANTS AND CONDENSATES FROM THE $V-A$ SPECTRUM}

\author{DIOGO BOITO}

\address{Physik Department T31, Technische Universit\"{a}t M\"{u}nchen\\ James-Franck-Stra{\ss}e 1, D-85748 Garching, Germany}

\author{MAARTEN GOLTERMAN }

\address{Department of Physics and Astronomy, San Francisco State University\\
San Francisco, CA 94132, USA}

\author{MATTHIAS JAMIN}

\address{Instituci\'{o} Catalana de Recerca i Estudis Avan\c{c}ats (ICREA),\\
IFAE,  Universitat Aut\`{o}noma de Barcelona, E-08193 Bellaterra, Barcelona, Spain}

\author{KIM MALTMAN}

\address{Department of Mathematics and Statistics, York University Toronto\\
Ontario M3J 1P3, Canada\\ and\\ CSSM, University of Adelaide, Adelaide\\
South Australia 5005, Australia}

\author{SANTIAGO PERIS\footnote{Speaker at the conference}}

\address{Department of Physics, Universitat Aut\`{o}noma de Barcelona\\
E-08193 Bellaterra, Barcelona, Spain}

\maketitle

\begin{history}
\received{Day Month Year}
\revised{Day Month Year}
\end{history}

\begin{abstract}
We present an analysis of the isospin-one $V-A$ correlator based on our successful simultaneous description of the OPAL $V$ and $A$ non-strange tau spectral data. We discuss the values obtained for the Chiral Perturbation Theory low-energy constants $L_{10}$ and $C_{87}$ as well as the dimension-six and eight condensates and compare them with those in the  literature.
\end{abstract}
\ccode{TUM-HEP-915/13}

Low-energy constants (LECs) and condensates are effective parameters in QCD encoding important non-perturbative information. While the former are key ingredients for a complete and systematic description of low-energy physics in Chiral Perturbation Theory (ChPT), the latter play an equally important role at higher energies where the Operator Product Expansion (OPE) becomes applicable. In the following, I will summarize a recent determination of these parameters which have appeared in Ref.~\refcite{us}.

Let us define the function
\begin{equation}\label{def}
\widehat{\Pi}^{(w)}_{V-A}(Q^2)=\int_0^\infty dt\ w(t/s_0)\ \frac{\rho_V(t)-\rho_A(t)}{t+Q^2}\ ,
\end{equation}
where $w(x)$ is a polynomial, $s_0$ is a parameter that will be conveniently chosen below, and  $\rho_{V,A}$ are the non-strange $I=1$, $J=0+1$ vector and axial-vector spectral functions \emph{without} the pion pole. For instance, the function $\widehat{\Pi}^{(1)}_{V-A}(Q^2)$ is nothing but the usual $\langle VV-AA\rangle$ correlator without the pion contribution. The goal is to calculate the coefficients $ L_{10}^{eff}$ and $C_{87}^{eff}$, related to the corresponding  $\mathcal{O}(p^4)$ and $\mathcal{O}(p^6)$ LECs of Chiral Perturbation Theory, appearing in the low-$Q^2$ expansion\cite{DGHS,ABT,GPP}
\begin{equation}\label{chpt}
\widehat{\Pi}^{(1)}_{V-A}(Q^2)= - 8\  {L_{10}^{eff}} - 16\ { C_{87}^{eff}}\ Q^2 + \cdots \ , \quad Q^2\rightarrow 0
\end{equation}
as well as the dimension-6 and -8 condensates $C_{(6,8);V-A}$ appearing in the OPE:
\begin{equation}\label{ope}
    \Pi^{OPE}_{V-A}(Q^2)= \frac{ C_{2,V-A}}{Q^2} + \frac{ C_{4,V-A}}{Q^4} + \frac{C_{6,V-A}}{Q^6} +  \frac{ C_{8,V-A}}{Q^8} + \cdots \ , \quad  Q^2\rightarrow \infty\ .
\end{equation}
Note that, unlike Eq. (\ref{chpt}), the OPE in Eq. (\ref{ope}) involves the full $\Pi_{V-A}(Q^2)$ i.e., it includes the pion pole. In Eq. (\ref{ope}), the coefficients $C_{(2,4); V-A}$ are known, and are proportional to $(\alpha_s m_q^2, \alpha_s m_{\pi}^2)$, respectively.\cite{Floratos,Chetyrkin} The main difficulty with the evaluation of the integral in Eq. (\ref{def}), and consequently with the determination of the LECs and condensates in Eqs. (\ref{chpt}) and (\ref{ope}), is the fact that the spectral data  stops at the tau mass and does not extend all the way up to infinity. An extrapolation function is needed.

\begin{figure}[h]
\centerline{\includegraphics[width=2in]{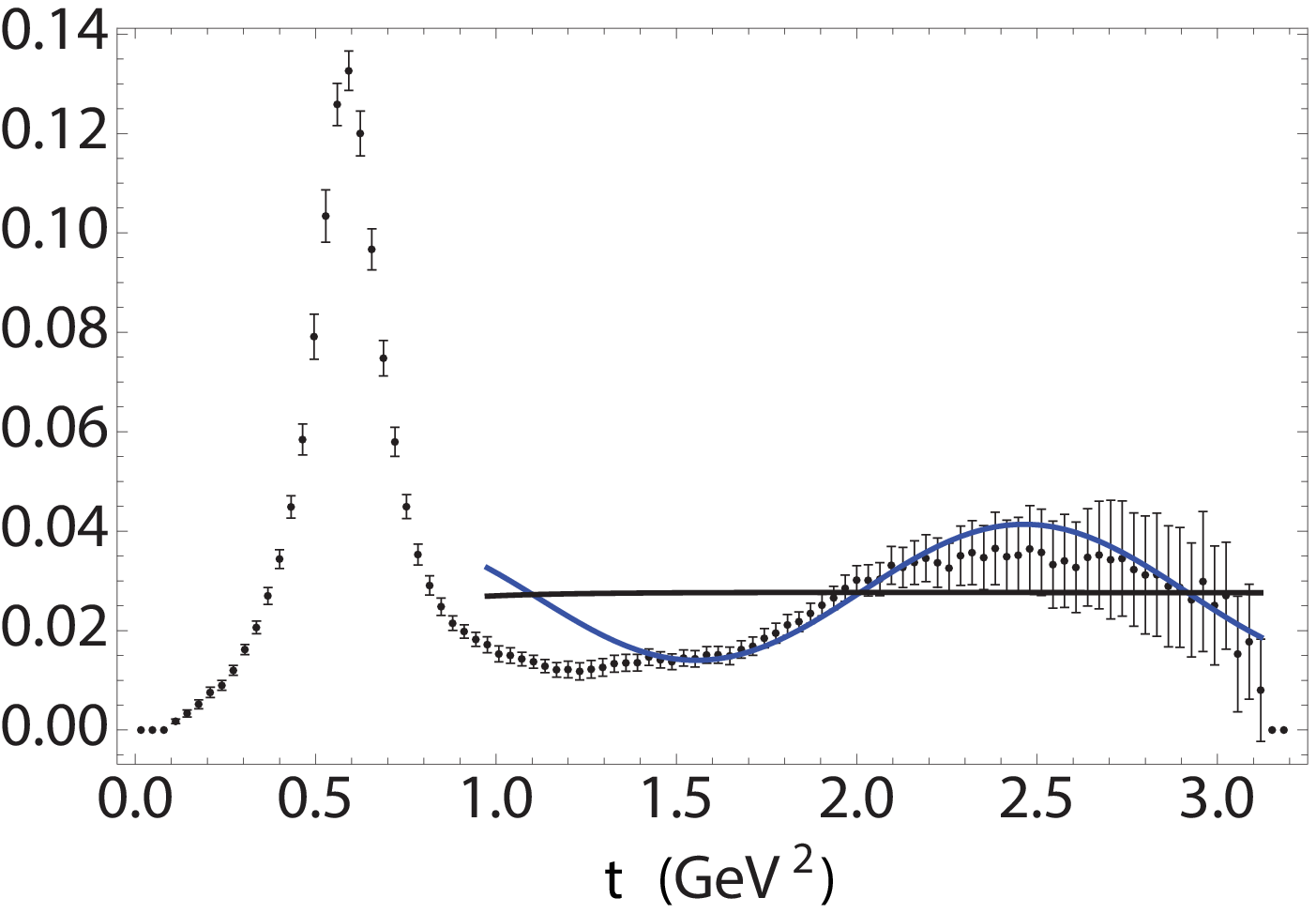}\hspace{.5cm}
\includegraphics[width=2in]{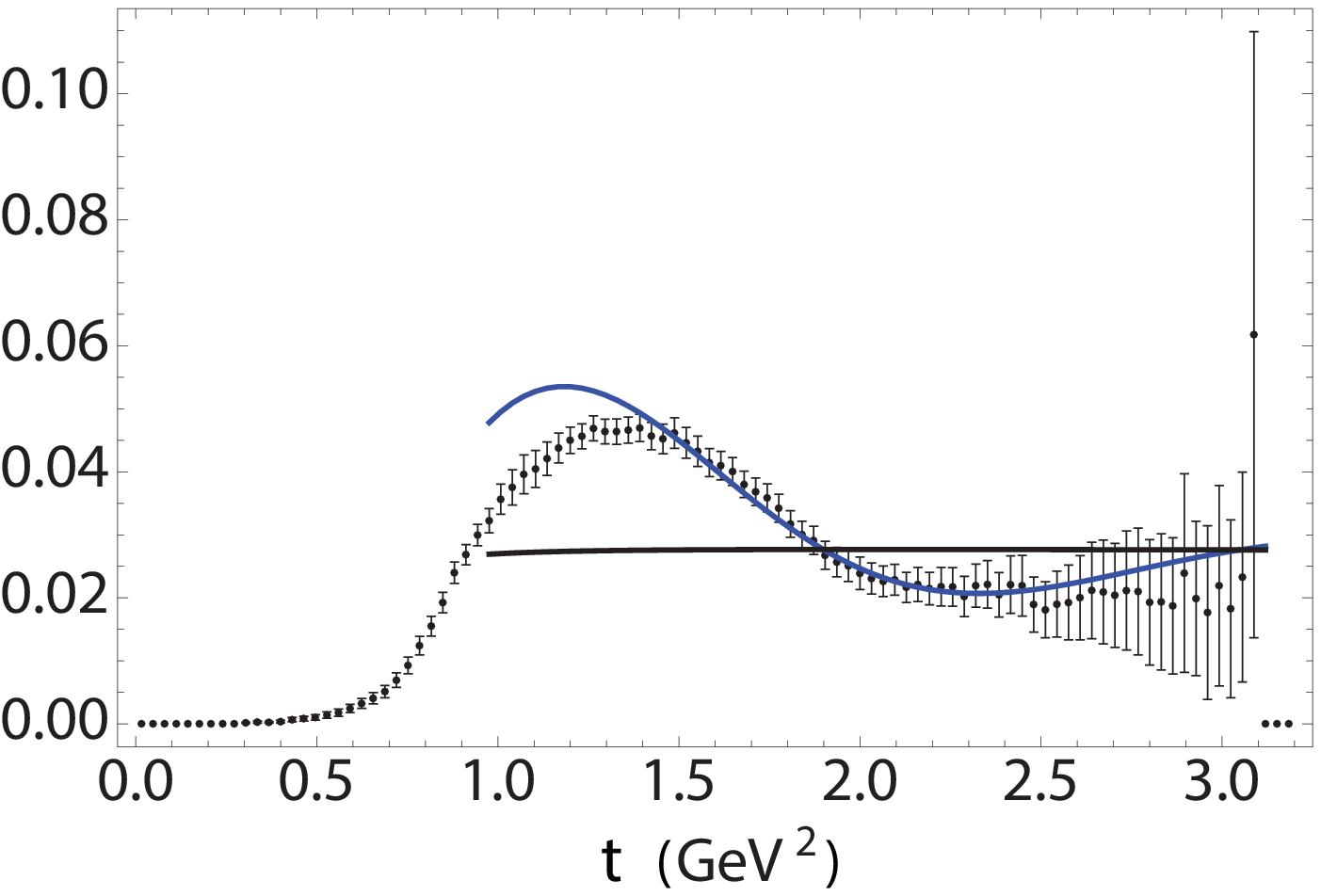}}
\vspace*{8pt}
\caption{OPAL $V$ and $A$ spectra compared to perturbation theory to order $\alpha_s^4$  (condensates contribute a negligible amount) plus the DV model, Eq. (\ref{rhodv}). The flat horizontal line corresponds to perturbation theory only, without the DV term. \label{Fig1}}
\end{figure}

Fortunately, analyticity constrains somewhat this extrapolation. When $s_0$ is large enough, the analytical properties of the two-point function $ \Pi_{V-A}$ enforce the constraint\cite{Floratos,DV}
\begin{eqnarray}\label{fesr}
     \int_{4m_{\pi}^2}^{s_0}dt\ w(t)\ \rho_{V-A}^{{Exp.}}(t) - 2\ f_{\pi}^2\ w(m^2_{\pi})+ \int_{s_0}^{{\infty}}dt\ w(t)\ \rho_{V-A}^{{ DV}}(t) & &  \\
  &&\hspace{-2.5cm} =  -\ \frac{1}{2\pi i}\oint_{|z|=s_0}dz\ w(z)\ \Pi^{{OPE}}_{V-A}(z) \nonumber
\end{eqnarray}
where we have split the correlator into its $OPE$ and Duality Violation ($DV$) parts i.e., $\Pi_{V-A}= \Pi^{OPE}_{V-A}+ \Pi^{DV}_{V-A}$  with $\rho_{V-A}^{{ DV}}= \frac{1}{\pi}\mathrm{Im}\,\Pi_{V-A}^{DV}$. The celebrated Weinberg sum rules\cite{weinberg} result from  Eq. (\ref{fesr}) when the polynomials $w(t)=1, t$ are chosen and the limit $s_0\rightarrow \infty$ is taken. For the DV part we will use the parametrization
\begin{equation}\label{rhodv}
    \rho^{{DV}}_{V/A}(t)=\ \mathrm{e}^{-\delta_{V/A} -\gamma_{V/A}t}\ \sin\left( \alpha_{V/A}+ \beta_{V/A} t\right)\quad ,
\end{equation}
for $t \geq s_0 \simeq 1.5\ \mathrm{GeV}^2$. For a discussion of the rationale behind this parametrization see Ref. \refcite{DV}, which is based on earlier studies in Ref. \refcite{DVbefore}.
In Fig. \ref{Fig1} we show how well this expression plus the perturbative spectral function to order $\alpha_s^4$ describes \emph{both} the $V$ and $A$ OPAL spectra. It is conceptually important that both channels are independently described, as opposed to some ``effective" description of the $V-A$ combination. DVs reflect the failure of the OPE to describe the correlator for  Minkowski momenta where resonances exist. To minimize model dependence, therefore, it is necessary to allow for the possibility of independent parametrizations of DVs for different spectra (i.e., different  quantum numbers) and treat $V$ and $A$ separately. In practice, the values of the parameters $\delta_{V/A} , \gamma_{V/A}, \alpha_{V/A}$ and $\beta_{V/A}$ employed can be found in Refs. \refcite{us,US2,US1}.

Once a good description of the spectral data is obtained, as in Fig. \ref{Fig1}, one may consider doing the integral in Eq. (\ref{def}) in the interval $4 m_{\pi}^2\leq t\leq s_0$ using the experimental data points, and the DV parametrization (\ref{rhodv}) for $s_0\leq t < \infty$. In practice the value $s_0\simeq 1.5\ \mathrm{GeV}^2$ produces both good and stable results and this is the one we have used. Choosing the weights $w_k(x)=(1-x)^k$ in Eq. (\ref{def}) one obtains
\begin{eqnarray}\label{results1}
   {L_{10}^{eff}} &=& -\frac{1}{8}\left(\widehat{\Pi}^{w_2}_{V-A}(0)+ \frac{4 f_{\pi}^2}{s_0}\left[ 1- \frac{m_{\pi}^2}{2 s_0}+ ...\right]\right) = -6.45(9)\times 10^{-3}\ ,\\
  C_{87}^{eff} &=& -\frac{1}{16} \frac{d}{dQ^2} \widehat{\Pi}^{w_0}_{V-A}(0)= 8.47(29)\times 10^{-3}\ \mathrm{GeV}^{-2}\nonumber .
\end{eqnarray}

Factors of $f_{\pi}^2$ and $m_{\pi}^2$ in Eq. (\ref{results1}) appear as a consequence of the first and second Weinberg sum rules, whereas the ellipses stand for terms which can safely be neglected. The values obtained for $L_{10}^{eff}$ and $  C_{87}^{eff}$ are compatible with those in Ref. \refcite{GPP2}, but our errors are $\sim 2$ larger. One reason for this is that  OPAL's errors are larger than ALEPH's data used by Ref. \refcite{GPP2}. However, there is a reason why we have only used OPAL data. This is because ALEPH's is currently flawed by a problem in its covariance matrices, as was first pointed out in Ref. \refcite{TAU10} and recently acknowledged in Ref. \refcite{MALAESCU}. It is unknown how much this flaw may affect the results. Once ALEPH's data is properly fixed, we may use it as well. The second reason why our errors are larger  is the oversimplified form for $\rho_{V-A}^{{ DV}}$ used by Ref. \refcite{GPP2}, which is like that shown in Eq. (\ref{rhodv}) but with only 4 parameters, and incompatible with the spectrum observed in the $V$ and $A$ separate channels.\cite{US2,US3} The systematic error associated with this choice is not included in the total error quoted in Ref. \refcite{GPP2}. We emphasize that $4(V)+4(A)=8$ parameters are needed in our case for a good description of both the $V$ and $A$ data.

Note that the results quoted in Eqs. (\ref{results1}) are just some effective parameters (cfr. Eq. (\ref{chpt})). Their  relationship with the true LECs of the ChPT Lagrangian $L^r_{10}(\mu)$ and  $C^r_{87}(\mu)$ is highly nontrivial. For example, for the case of  $L^r_{10}(\mu)$, this relationship depends on the contribution from terms which are $\mathcal{O}(p^6)$ (and higher) in the ChPT expansion. These $\mathcal{O}(p^6)$ terms already involve other LECs, which are sometimes not very well-known. Therefore, to extract a value for $L^r_{10}(\mu)$ one needs to decide what to do with these unknown $\mathcal{O}(p^6)$ LECs. To resolve this issue, Ref. \refcite{GPP2} was forced to make some model assumptions based on Vector meson Dominance and large-$N_c$-inspired arguments which, although perhaps reasonable for an order-of-magnitude estimate, are not sufficiently robust to bring the systematic error under good theoretical control. This is why in Ref. \refcite {us} we decided to follow a different route and take  advantage of the possibility that exists on the lattice to vary the quark masses. Using this trick, lattice data\cite{Boyle} and our continuum constraints, it was possible  to determine with sufficient accuracy these unknown $\mathcal{O}(p^6)$ LECs, leading to the value
\begin{equation}\label{L10}
    L^r_{10}(M_{\rho})=-3.1(8)\times 10^{-3},
\end{equation}
with again a factor of $\sim 2$ larger errors than those in Ref. \refcite{GPP2}. In fact, the lattice data violates some of the model assumptions made in Ref. \refcite{GPP2}. A recent analysis\cite{Kim} of combined lattice and continuum data, using the flavor breaking combination $ud-us$ of the chirally breaking combination $\langle VV-AA\rangle$, obtained the result
\begin{equation}\label{L10Kim}
    L^r_{10}(M_{\rho})=-3.46(29)\times 10^{-3},
\end{equation}
in very good agreement with (\ref{L10}).

Analogously to the case of $ L^r_{10}(M_{\rho})$, also $C^{eff}_{87}(\mu)$ receives contributions from LECs (in this case at $\mathcal{O}(p^8)$). These contributions, however, have not been calculated in ChPT. In this case, we can only make a very rough guess and estimate the error as $\sim 25\%$ (e.g. the typical size of missing chiral corrections\footnote{This typical size of $\sim 25\%$ in the chiral corrections were present in the case of the determination of $L^r_{10}(M_{\rho})$ in Eq. (\ref{L10}). Furthermore, an exploration of the convergence of ChPT for the correlator (\ref{def}) also confirms the presence of these corrections.}) in the value inferred for $C^r_{87}(M_{\rho})$, which turns out to be
\begin{equation}\label{C87}
   C^r_{87}(M_{\rho})= 4(1)\times 10^{-3}\ \mathrm{GeV}^{-2} .
\end{equation}

The problem of obtaining the OPE condensates $C_{(6,8);V-A}$ in Eq. (\ref{ope}) is even harder. Because these condensates appear in the large-$Q^2$ expansion, they are determined by positive moments of the spectral function $\rho_{V-A}(t)$ with the powers $t^2$ and $t^3$, respectively. This results in a strong sensitivity to details in the upper end of the physical spectrum (where the data points have larger errors) as well as to the inclusion of DVs (with the consequent increase in model dependence).  However, it is important to realize that neglecting DVs altogether, as it is sometimes done, is as much a model as our ansatz in Eq. (\ref{rhodv}). It corresponds to the choice $\delta_{V/A}\rightarrow \infty$ which, in light of Fig. \ref{Fig1}, is not a particularly good model.  The choice of polynomials\cite{Yavin,GPP2} $w(t)=(t-s_0)^2$ and $w(t)=(t-s_0)^2 (t+2 s_0)$ ameliorates the situation somewhat, suppressing at the same time both the contribution from the higher end of the spectrum as well as the one from DVs,  and effectively replacing their contribution, with the help of the Weinberg sum rules, by pion-pole terms. Our results are the following:
\begin{eqnarray}
\label{C6C8t3}
C_{6,V-A}&=&(-6.6\pm 1.1)\times 10^{-3}\ \mbox{GeV}^6
\ ,\nonumber\\
C_{8,V-A}&=&(5\pm 5)\times 10^{-3}\ \mbox{GeV}^8\
\ .
\end{eqnarray}

We refer the reader to Ref. \refcite{us} for more details as well as a comparison with the different results found in the literature for these condensates (see Fig. 3 in this reference).
\section*{Acknowledgments}

DB is supported by the Alexander von Humboldt Foundation,
MG is supported by the US DOE, MJ and SP are supported by CICYTFEDER-FPA2011-25948, SGR2009-894, CPAN (CSD2007-00042) and KM is supported by NSERC (Canada).


\end{document}